# Hexagonal boron nitride thin film synthesis with a ns–pulsed MHCD: *in–situ* plasma diagnostics and post–growth film characterization

Belkacem Menacer, Dimitrios Stefas, Nikolaos Chazapis,
Claudia Lazzaroni[a], Kristaq Gazeli[a], and Vianney Mille

Laboratoire des Sciences des Procédés et des Matériaux (LSPM – CNRS), Université Sorbonne Paris Nord, Villetaneuse, UPR 3407, F-93430, France

[a]emails: claudia.lazzaroni@lspm.cnrs.fr; kristaq.gazeli@lspm.cnrs.fr

## ABSTRACT

Hexagonal boron nitride (h-BN) is deposited on Si <100> wafer (≈20 cm$^2$) via Plasma Enhanced Chemical Vapor Deposition (PECVD) using a ns-pulsed $N_2$/Ar Micro Hollow Cathode Discharge (MHCD) as a microplasma source. For the first time, aluminum nitride (AlN) is employed as the dielectric material in the MHCD to mitigate film contamination by atomic oxygen, an issue previously observed with conventional $Al_2O_3$ dielectrics. A comprehensive multi-diagnostic approach is followed to characterize the deposited h-BN, including Raman spectroscopy, scanning electron microscopy (SEM), atomic force microscopy (AFM), and X-ray photoelectron microscopy (XPS). In parallel, *in-situ* diagnostics such as optical emission spectroscopy (OES) and intensified CCD (ICCD) imaging are used to monitor plasma properties, including emission profiles, gas temperature and discharge morphology. Raman spectra reveal the $E_{2g}$ phonon mode of h-BN around 1366 cm$^{-1}$, confirming successful synthesis. SEM imaging reveals an almost complete surface coverage by the film, with localized delamination. This is probably due to an uneven resistive heating of the Si wafer, rapid post-deposition cooling (∼13 K/min) and ambient exposure. AFM analysis indicates an average thickness of about 33 nm after 90 minutes of deposition (∼22 nm/h deposition rate). XPS measurements reveal an average B/N atomic ratio of ∼1.5 along the wafer diameter. Deviations from ideal film properties (e.g., stoichiometric unity, uniform morphology) are attributed to plasma-induced inhomogeneities (such as non-uniform species flux and temperature gradients) among other factors (e.g., ambient exposure post-deposition), which affect nitrogen and boron incorporation and localized film properties. Despite these challenges, the MHCD-driven PECVD process demonstrates strong potential for scalable h-BN synthesis, with further optimization of the reactor design, plasma conditions, and gas chemistry required to grow ideal films.

## I. INTRODUCTION

Plasma Enhanced Chemical Vapour Deposition (PECVD) allows for the synthesis of thin films of different materials at lower deposition temperatures ($T_D$) compared to conventional processes[1]. For instance, PECVD driven by ns-pulsed Micro Hollow Cathode Discharges (MHCDs) has been proposed to synthesize hexagonal boron nitride (h-BN) at $T_D \leq 1073$ K[2,3]. Also known as white graphite, h-BN is a layered material with a large bandgap (about 6 eV), remarkable thermal conductivity, chemical inertness, breakdown voltage, mechanical strength, etc.[4]. These properties make it a key material for a wide range of applications in electronics, spintronics, and optoelectronics[5,6]. For instance, h-BN can be combined with graphene (Gr) to form vertical Gr/h-BN/Gr heterostructures[5]. These can be used to construct new field effect transistors (FET) by overcoming the current limitations of the traditional FET technologies [5]. Thus, the development of new reliable PECVD processes for sustainable synthesis of h-BN can be a step toward advancing next-generation devices based on 2D materials. Toward this direction, MHCD technology offers key advantages such as low power consumption, high electron density, rich chemical reactivity and relatively low gas-phase temperature (typically <800 K[7–11]). These features make it suitable for an effective dissociation of molecular nitrogen (abundant user-friendly precursor of N–atoms) which is generally difficult to dissociate[10,12].

A typical MHCD setup consists of a circular sandwich structure comprising two molybdenum electrodes separated by an alumina (Al$_2$O$_3$) dielectric[10,13–15]. The ensemble is glued together and a micrometric hole (~100s μm in diameter) is drilled at the center of the assembly. The sandwich is placed in the junction between a lower-pressure (several mbar) and a higher-pressure (several tens mbar) chamber. The discharge can be generated in different gases (N$_2$, N$_2$/Ar, He, …) using DC or pulsed high voltages (HV)[2,7–9,16–18].

The pressure difference is essential in ns-pulsed MHCDs used for material synthesis for the transport of essential atoms (such as N–atoms in the case of h-BN synthesis) to a substrate placed on a heated and polarisable holder. Such an MHCD, using Al$_2$O$_3$ dielectric and generated in N$_2$/Ar (mixed with a boron precursor, BBr$_3$), was used for the synthesis of h-BN films over a 5-cm-diameter Si <100> wafer[2]. Film characterization was done using Raman and electron energy loss spectroscopies (EELS) as well as transmission electron microscopy (TEM). Despite the feasibility of h-BN deposition with MHCD, this study revealed the need of improved investigations towards obtaining a better quality of the h-BN films synthesized. Specifically, the film presented heterogeneities in thickness and color, holes and weak adherence on the Si substrate. Another issue referred to the presence of atomic oxygen in the films. Besides other factors (e.g., film contact with ambient air after deposition), this issue may also originate in the dielectric material of the MHCD chosen (Al$_2$O$_3$). Degradation of the dielectric material by the energetic discharge in the MHCD hole is possible since the power density in the hole can reach large values (~kW cm$^{-3}$). Another cause could be due to the presence of impurities in the gases and precursors used to form h-BN. This underlines the need of exploring new constructive materials towards improving the MHCD technologies for h-BN deposition.

This work is a first investigation of a ns-pulsed N$_2$/Ar MHCD using aluminum nitride (AlN) as dielectric material instead of classic Al$_2$O$_3$[2,3,7–10,13–18]. The aim is to demonstrate the reliable operation of this MHCD with AlN and interrogate its efficiency to synthesize h-BN films on large-area substrates (20 cm$^2$), while limiting (as much as possible) the presence of O–atoms in the films. Furthermore, contrary to previous works where only the h-BN properties were considered post-deposition by analyzing them via Raman, EELS and TEM[2], and by Raman, AFM and electron probe micro-analysis (EPMA)[3], this study represents a multi-diagnostic approach. The plasma features are mapped *in-situ* during deposition to make a correlation with the film properties measured post-deposition.

The paper is organised as follows. Section II is devoted to the experimental setup and methodology. Section III presents the results obtained: (i) plasma features via electrical measurements, optical emission spectroscopy (OES) and intensified CCD imaging, (ii) film properties via Raman, SEM, AFM and XPS. The main conclusions and perspectives are presented in section IV.

## II. EXPERIMENTAL SETUP AND METHODOLOGY

The MHCD studied in this work consists of two circular molybdenum electrodes (100 μm thickness) and an AlN plate as dielectric (750 μm thick), as shown in **Fig. 1(a)**. The MHCD hole has a diameter of 250 μm, being 1.6 times smaller than in previous works[2,3]. A 3D representation of a section of the deposition reactor is shown in **Fig. 1(b)**. The high-pressure chamber is maintained at $P_{high}$=30 mbar and the low-pressure (deposition) chamber at $P_{low}$=1 mbar, the pressure difference being about 1.7 and 2 times smaller than in ref.[2] (35/0.7 mbar) and in ref.[3] (45/0.75 mbar), respectively. The discharge operates in a N$_2$/Ar gas mixture (1:1 ratio; 99.999% purity; 80 sccm total flow rate; O$_2$<2 ppm, H$_2$O<3 ppm) and is powered by a negative ns-pulsed power supply (–1 kV peak, 500 ns width, 10 kHz frequency; a typical voltage pulse is shown in **Fig. 2**). The gas mixture is introduced into the high-pressure chamber and flows into the low-pressure chamber through the MHCD hole. The boron precursor, boron tribromide (BBr$_3$; 99.99% purity), is directly injected (via argon bubbling) into the deposition chamber (see **Fig. 1(b)**). Besides, we also injected H$_2$ gas (99.999% purity; O$_2$≤2 ppm, H$_2$O≤3 ppm, CO≤0.5 ppm, CO$_2$≤0.5 ppm) to limit amorphous BN synthesis, as in ref.[3]. The flow rates of BBr$_3$ and H$_2$ are set to 10 and 25 sccm, respectively.

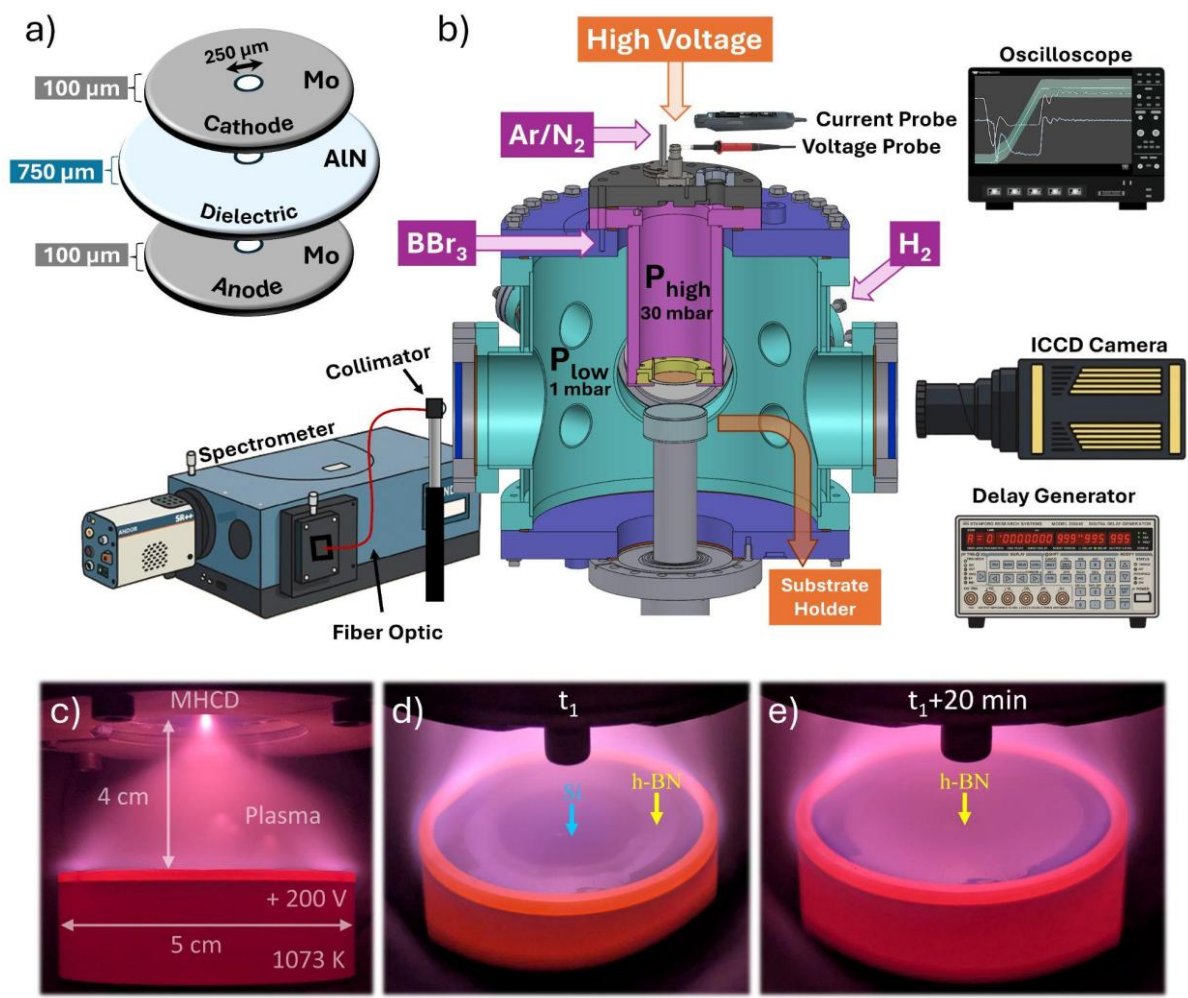

**FIG. 1**. (a) MHCD assembly used in this work, (b) experimental setup of the deposition reactor and different plasma diagnostics, (c) plasma expansion in the low-pressure chamber during h-BN deposition on a Si wafer placed on a heated (1073 K) and polarizable (+200 V) holder, (d),(e) views of the h-BN film deposited in two different instants ((d): t$_1$ and (e): t$_1$+20 min) from the beginning of the deposition. Photos (c)–(e) are captured with a conventional camera with the following settings: f/1.9, 50 ms exposure time (500 HV pulses integration), ISO200.

A holder is placed at a distance d=4 cm (≈1.4 times smaller than in ref.[2] and close to that in ref.[3]) from the exit of the MHCD's hole. This is biased with a DC voltage (+200 V) to extend the plasma toward a 5-cm-diameter Si <100> substrate placed on it. Si is a widely used substrate in PECVD, and is adopted here due to its smooth surface, thermal stability, and compatibility with h-BN. Si <100> was also used in ref.[2] as h-BN substrate, while deposition of h-BN on Si <111> and sapphire <0001> was reported in another work[3]. Furthermore, to improve the film quality, h-BN adhesion on substrate, and overall deposition control, the holder is heated to 1073 K via a graphite resistor (**Fig. 1(c)**). This also helps in enabling boron atom formation and diffusion towards the substrate, facilitating h-BN formation. The h-BN synthesis is performed for 90 minutes, which is 3 times larger than in ref.[3], and 1.3 and 4.6 times smaller than the deposition times used in ref.[2]. **Fig. 1(c)** depicts an indicative time-integrated photograph of the plasma expansion in the low-pressure chamber during the synthesis of h-BN. The formation of the h-BN film in two indicative instants during the deposition process is shown in **Fig. 1(d)** and **Fig. 1(e)**, respectively. Once the deposition is finished, the biasing and heating of the substrate holder are stopped, and it cools down to ambient temperature (cooling rate: 13 K/min).

Then, the substrate with the film is removed from the deposition chamber for subsequent analysis. The films are characterised using Raman spectroscopy (Jobin-Yvon HR800 with 473 nm laser; 1 μm spot size) for structural analysis, SEM (Zeiss MEB FEG SUPRA 40VP) for surface morphology, AFM (Bruker Dimension Icon ScanAsyst; using aluminum lacquer as conductive coating) for thickness and surface roughness, and XPS (K-Alpha+, ThermoFisher Scientific) for elemental composition and surface contamination. Film analyses are performed with the following order: Raman performed about 1 h after the end of their synthesis, SEM 3 days after, AFM 5 days after, and XPS 2 weeks after. The delays for performing SEM, AFM and XPS were due to the unavailability of each diagnostic for immediate use.

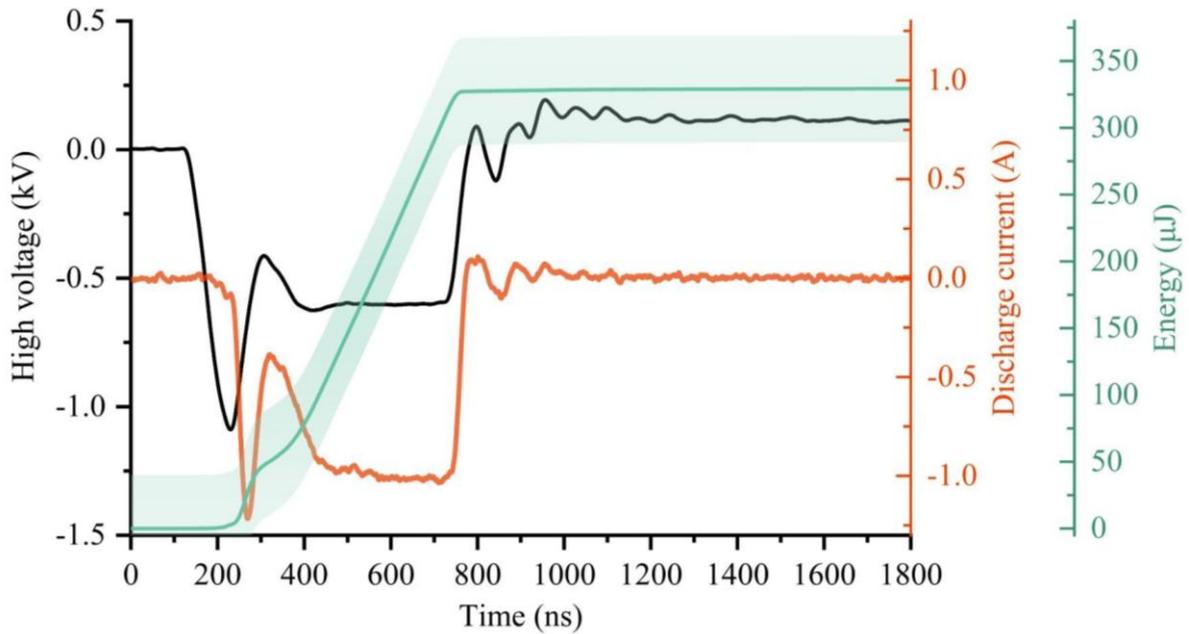

**FIG. 2**. Indicative applied voltage, induced current, and energy deposited in the discharge (average value from 10 consecutive measurements taken during the synthesis of two different films).

For studying the plasma features, the applied voltage (HV) to the MHCD electrodes is measured using an HV passive probe (LeCroy PPE 6 kV, 100:1, 400 MHz) and the total current is monitored by means of a current probe (LeCroy CP030, 50 MHz) connected to a digital oscilloscope (LeCroy HDO6104; 1 GHz, 2.5 GS/s). Then, the discharge current ($I_{dis}$) is obtained by subtracting the capacitive current ($I_C$=C×dV/dt, C being the equivalent capacitance of the MHCD, here estimated to be around 45 pF) from the total current, as in [19]. Indicative electrical signals of the applied voltage and the discharge current are depicted in **Fig. 2**. These waveforms are used to calculate the energy deposited in the discharge ($E_{dis}$=330±40 μJ) at the end of the voltage pulse, as follows:

$$E_{dis} = \int V(t) \times I_{dis}(t) dt \quad (1)$$

The space-integrated discharge's emission in the deposition chamber is collected using a collimating lens (Avantes COL-UV/VIS; 200–2500 nm) and guided using an optical fiber (Avantes FC-UVIR200-1) to the entrance slit (10 μm width) of a portable spectrometer (Avantes AvaSpec-ULS4096CL-EVO; 75 mm focal length). The latter has a 300 lines/mm grating and a linear CMOS sensor (2048 px) covering a spectral range from 200 to 1100 nm with a 0.7 nm resolution. The emitted light is recorded using 5 s integration time (corresponding to $5\times10^3$ HV pulses). This system mainly allows to identify the emissive reactive species generated and assess their formation mechanisms. Furthermore, the emission pattern in the deposition chamber is captured using an ICCD camera (PIMAX-1K-RB-FG-P43; Princeton Instruments) coupled to a UV lens. In this case the emitted light is recorded using a gate width of 100 µs and accumulated over 100 HV pulses. Finally, we use a 500-mm-focal-length spectrograph (Andor Shamrock 500i; 2400 l/mm ruled grating blazed at 300 nm; 50 µm slit; 0.05 nm resolution) backed with an ICCD camera (Andor iStar sCMOS 18U-03). This allows capturing space-integrated rotational spectra (500 ns gate width; $10^5$ accumulations) of $N_2$(SPS) and estimating the gas temperature through the rotational temperature of $N_2$(C)[7,8,20]. The synchronisation of the different units is done using a delay generator (DG645; Stanford Research Systems). The detailed experimental setup is presented in **Fig. 1**.

### III. RESULTS AND DISCUSSION

The choice of the MHCD as an atomic nitrogen source is done because of its demonstrated efficiency in dissociating molecular nitrogen[9,10,12]. Particularly, it was shown that DC MHCDs enable the generation of high electron densities ($N_e$) which can induce various relevant reactions driving the plasma chemistry and reactivity (e.g., formation of excited nitrogen and Ar including $N_2$(A) and Ar metastables)[9,11]. In a DC MHCD, $N_e$ was determined to be on the order of $10^{14}$ cm$^{-3}$ [9]. Interestingly, when operated in the self-pulsing mode, $N_e$ reached peak values up to $4\times10^{15}$ cm$^{-3}$ [11] which may be useful for the $N_2$ dissociation. However, self-pulsing MHCDs are prone to intrinsic instabilities which affect the reproducible production of various plasma components. Fortunately, using ns-pulsed HV to drive MHCDs is a convenient solution to generate high electron densities, while achieving reproducible plasma features and also avoiding arcing because of the small pulse durations applied[7,8].

Furthermore, the use of an $N_2$/Ar gas mixture is beneficial for an easier dissociation of $N_2$ toward h-BN formation [2,3,9,10]. This is possible because of an efficient energy transfer from argon metastables to nitrogen molecules leading to their excitation to high ro-vibrational levels, thus facilitating their dissociation [9,12,21]. Understanding the role of metastables in the deposition process is essential for its optimization. However, their quantification in the deposition reactor using known laser-based methods (such as tunable laser absorption spectroscopy for Ar($1s_3$/$1s_5$) [9]) is challenging due to the complex setups of these diagnostics and the bulky size of the deposition reactor itself. In this case, "model" portable reactors are used to perform relevant fundamental studies [9]. For the deposition reactor, however, a more convenient solution which may provide indirect information of the plasma-induced chemistry is to use conventional optical diagnostics such as ICCD imaging and OES. These allow monitoring the discharge morphology in the deposition chamber and identifying important reactive species generated. Corresponding results obtained are shown in **Fig. 3**.

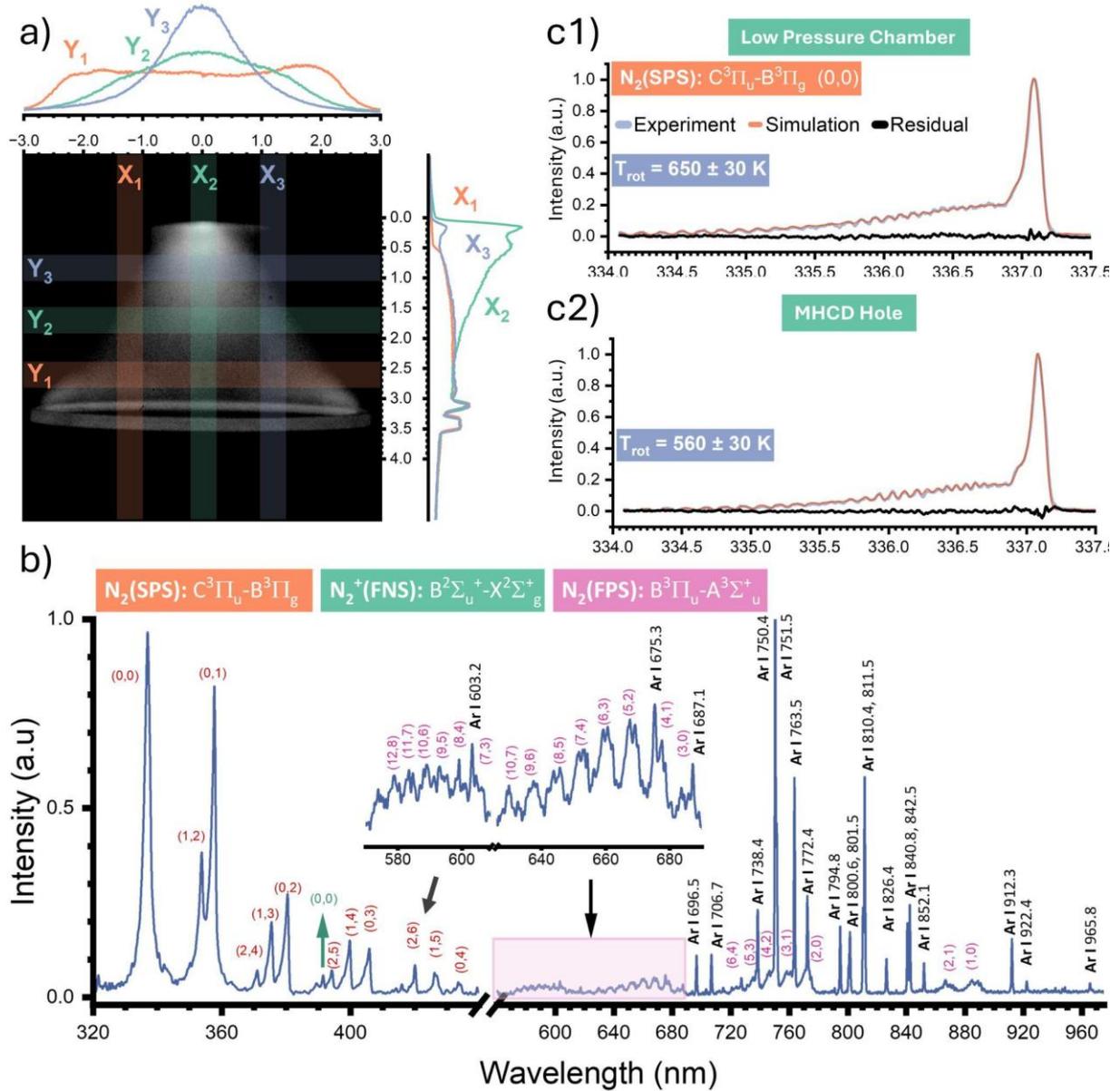

**FIG. 3.** (a) ICCD image (gate width=100 µs; 100 accumulations) of the wavelength-integrated emission in the low-pressure chamber; the intensity distribution along three representative horizontal ($Y_1$, $Y_2$, $Y_3$) and vertical ($X_1$, $X_2$, $X_3$) zones is shown. (b) Emissive reactive species generated in the low-pressure chamber. (c) Experimental rotational spectra of the $N_2$(SPS) in the low-pressure chamber (c1) and the MHCD's hole (c2) fitted with synthetic ones to estimate the corresponding rotational temperatures of $N_2$(C).

**Fig. 3(a)** shows a representative ICCD image with the emitted light accumulated over $10^3$ HV pulses, revealing a bell-shaped emission pattern. A similar profile has been obtained by capturing light over only one HV pulse, indicating a reproducible behavior during the deposition process. However, the spatial distribution of the emitted light along the height and width of the bell-shaped profile presents inhomogeneities. This is verified by plotting the variation of the intensity along three representative horizontal and vertical zones, their centers being at $Y_1$=2.75 cm, $Y_2$=1.75 cm, $Y_3$=0.75 cm, and at $X_1$=-1.25 cm, $X_2$=0 cm, $X_3$=1.25 cm, respectively. Analysis of the intensity's distribution along the $Y_{1,2,3}$ and $X_{1,2,3}$ zones reveals peak values on the center of the MHCD hole ($X_2$=0 cm) and far away from the substrate holder ($Y \geq Y_3$=0.75 cm). Besides, the intensity drops in the transversal ($X<X_2$ and $X>X_2$) and axial ($Y>Y_3$) directions. As expected, the discharge pattern shows an improved horizontal uniformity at $Y_2$ compared to that at $Y_3$. Then, it exhibits a quasi-uniform distribution close to the substrate surface ($Y_1$), where the peak intensity is almost the same between the center ($X_2$, $Y_1$) and the edges (($X_1$, $Y_1$) and ($X_3$, $Y_1$)) of the substrate. This, however, is significantly smaller than the peak intensity measured at ($X_2$, $Y_3$). Finally, the extracted profiles along $X_1$ and $X_3$ are slightly asymmetric for Y<0.5 cm, possibly due to an intrinsic asymmetry in the MHCD hole. Remigy *et al.* [10] used a DC MHCD with a polarised holder and performed similar spatial mappings of the absolute density of N–atoms generated in the low-pressure chamber of a portable "model" reactor. The peak N–atoms density was obtained on the MHCD axis (corresponding to $X_2$ in **Fig. 3(a)**) without observing any significant axial gradient between the MHCD's hole and the substrate. Furthermore, close to the substrate (corresponding to $Y_1$ in **Fig. 3(a)**) only a slight density decrease along its diameter was measured. However, in the bulk of the plasma a more abrupt density drop in the radial direction was revealed. An overall qualitative agreement is obtained between this work and ref.[10]. Any dissimilarities should be attributed to the different operating conditions used in ref.[10] such as DC-driven "model" MHCD, hole diameter (400 µm), bias voltage of the holder (+80 V), pressure in the deposition chamber (10 mbar), and gas mixture (80%$N_2$— 20%Ar). The spatial variations of the discharge components in the low-pressure chamber are probably representative of important hydrodynamic effects developed (e.g., supersonic gas flow velocities induced by the pressure difference between the high- and low-pressure chambers as well as the small size of the MHCD's hole[10]). These effects are expected to be more pronounced in the deposition reactor due to the smaller diameter of the MHCD hole (250 µm) compared to ref.[10], and are also likely to induce non-uniformities on the deposited films of h-BN, lowering their quality.

**Fig. 3(b)** illustrates a wide emission spectrum captured in the low-pressure chamber. This is space-integrated to improve the signal-to-noise ratio, which drops significantly due to the use of an optical fiber/collimator system and a narrow entrance slit on the spectrometer. A rich chemical reactivity is revealed in the range 320–975 nm (emissions below 320 nm cannot be recorded, being filtered by the optical window of the reactor). Specifically, we identify the following radiative transitions: $N_2$(SPS: $C^3\Pi_u$, v' → $B^3\Pi_g$, v") between 320 and 440 nm, $N_2$(FPS: $B^3\Pi_u$, v' → $A^3\Sigma_u^+$, v") between 560 and 900 nm, $N_2^+$(FNS: $B^2\Sigma_u^+$, v'=0 → $X^2\Sigma_g^+$, v"=0) at 391.4 nm, and various Ar I lines between 600 and 975 nm. This is the first real-time recording of different emissive species generated during h-BN deposition, some of them playing a role in the production of atomic nitrogen. For instance, an important mechanism responsible for the ro-vibrational excitation (and eventual dissociation) of $N_2$ is the energy transfer from argon metastables, as follows [9,21–25]:

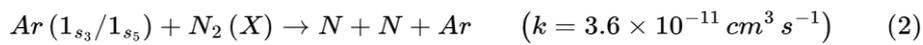
$$Ar\ (1_{s_3}/1_{s_5}) + N_2\ (X) \rightarrow N + N + Ar \qquad (k = 3.6 \times 10^{-11}\ cm^3\ s^{-1}) \qquad (2)$$

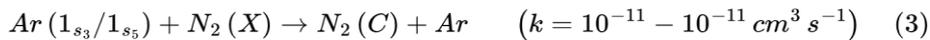
$$Ar\ (1_{s_3}/1_{s_5}) + N_2\ (X) \rightarrow N_2\ (C) + Ar \qquad (k = 10^{-11} - 10^{-11}\ cm^3\ s^{-1}) \qquad (3)$$

**Fig. 3(c)** depicts $N_2$(SPS) experimental rotational spectra recorded in the MHCD hole (**Fig. 3(c1)**) and in the low-pressure chamber (**Fig. 3(c2)**). These are fitted with synthetic ones using a custom-built code compared to massiveOES, showing a good agreement[26]. The rotational temperature ($T_{rot}$) of $N_2$(C) in the low-pressure chamber is around 650±30 K which is 110 K larger than the $T_{rot}$ in the MHCD hole. Considering this as an upper limit of the gas temperature, this value is about 1.5 times smaller than that of the holder and at least 2 times smaller than the values used by more conventional methods such as high temperature molecular beam epitaxy[1]. The difference in the $T_{rot}$ between the MHCD hole and the deposition chamber is due to an overpopulation of the $N_2$(C, v') in high rotational levels by Ar metastables (reaction (3))[21,22,24,25]. Their generation in the low-pressure chamber is uncovered by various Ar I lines in the spectrum, e.g., $2p_3$–$1s_5$ at 706.7 nm, $2p_6$–$1s_5$ at 763.5 nm, $2p_2$–$1s_3$ at 772.4 nm and $2p_4$–$1s_3$ at 794.8 nm. Furthermore, free electrons can be transported/generated in the deposition chamber due to the potential difference between the MHCD's anode (0 V) and the polarised holder (+200 V)[12]. These are also expected to be responsible for the formation of $N_2^+$($B^2\Sigma_u^+$, v'=0) state relaxing radiatively to the $N_2^+$($X^2\Sigma_g^+$, v"=0) state (i.e., $N_2^+$(FNS) transition at 391.4 nm). Finally, population to high vibrational states (up to v'=12) of $N_2(B)$ is revealed (see inset in **Fig. 3(b)**); this state de-excites to different vibrational levels (up to v"=8) of the $N_2(A)$ metastable state ($N_2$(FPS) transition in **Fig. 3(b)**) which may also contribute to N–atoms production, as follows[12,27]:

$$N_2(A) + N_2(A) \rightarrow N_2(X) + N + N \quad (k = 3 \times 10^{-11} \, cm^3 \, s^{-1}) \quad (4)$$

Nevertheless, emission from atomic nitrogen is not observed in the spectrum probably due to the following reasons: i) recombination between N–atoms and B–atoms forming h-BN, which is expected to be dominant in MHCDs used for h-BN deposition[2,3], ii) diffusion towards the reactor's walls[28,29], iii) recombination of N–atoms between them to form vibrationally excited $N_2(B)$ (evidenced by the $N_2$(FPS) bands detected in **Fig. 3(b)**), as follows[22,28–31]:

$$N(^4S) + N(^4S) + N_2 \rightarrow N_2(B, v') + N_2 \quad (k_1 = 8.27 \times 10^{-34} \exp(500/T(K)) \, cm^6 \, s^{-1}) \quad (5)$$

$$N(^4S) + N(^4S) + Ar \rightarrow N_2(B, v') + Ar \quad (k_2 = 2/6.5 \times k_1) \quad (6)$$

The above-discussed discharge properties define the spatial distribution of N-atoms in the low-pressure chamber and, consequently, the recombination efficiency with boron atoms and, eventually, the properties of the h-BN film deposited. These are studied in **Figs. 4–6**.

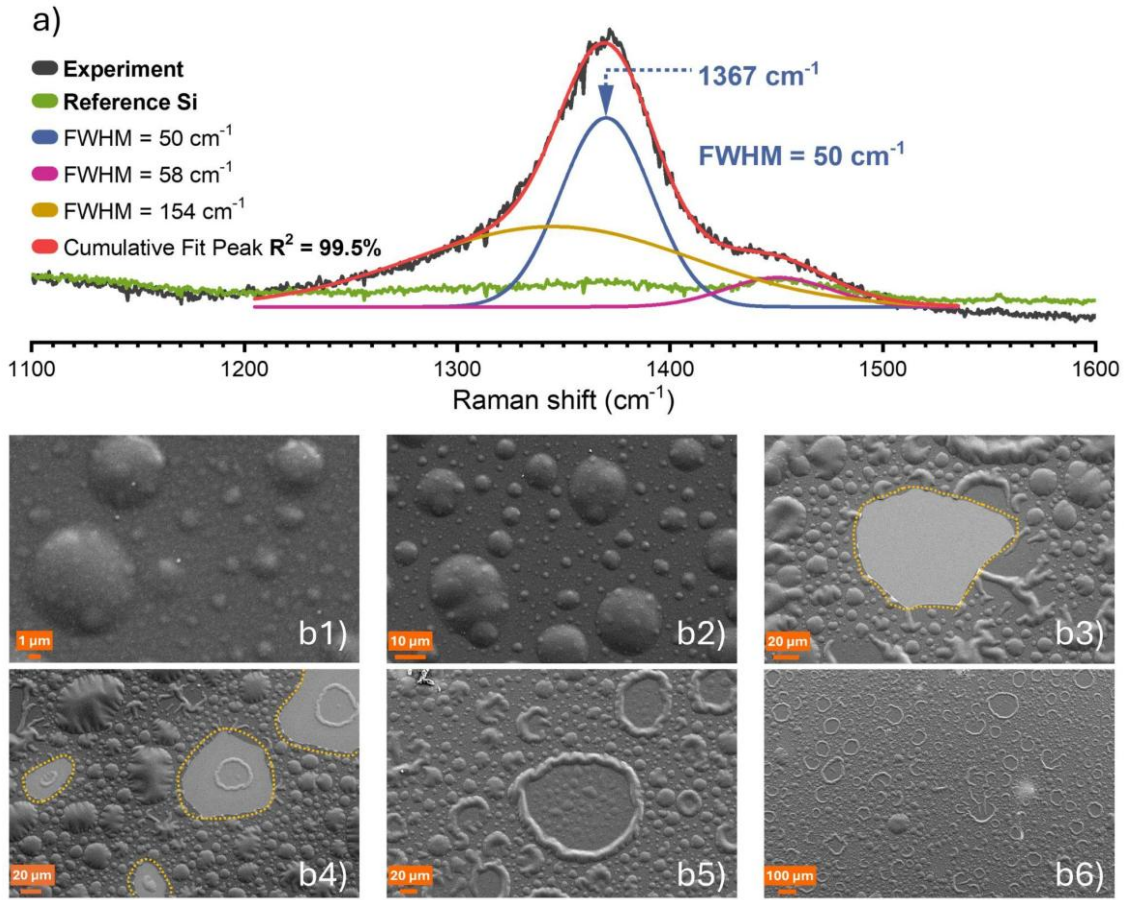

**FIG. 4**. (a) Experimental raw Raman spectrum (black) of a BN film deposited on a Si <100> substrate. The red line is a convolution of three Gaussians, used to achieve the best fit on the raw spectrum. This reveals the $E_{2g}$ signature of $sp^2$-h-BN around 1367 $cm^{-1}$ (blue Gaussian profile). The left wing of the raw profile is well-fitted with a broader Gaussian (yellow). The reference signal of Si is shown in green, its peak (around 1450 $cm^{-1}$) coinciding with that of a third Gaussian (magenta) used to construct the fitting line. (b) High-resolution SEM images of h-BN using different magnifications ((b1): 4000X; (b2): 1000X; (b3)/(b4)/(b5): 465X; (b6): 67X).

First, the deposited film is characterized using Raman spectroscopy directly after its extraction from the deposition chamber. The laser spot size used is about 1 μm allowing for a detailed analysis of the localized film properties (structure, bonding) with a decent spatial resolution. Numerous locations along the film surface are analysed showing consistent features. **Fig. 4(a)** shows a representative raw Raman spectrum which can be best fitted with a convoluted profile composed by different Gaussian functions, as in refs.[1,3]. The central Gaussian (blue line), peaking around 1367 cm$^{-1}$, represents the characteristic in-plane $E_{2g}$ vibrational mode of h-BN being a signature of sp$^2$-bonded hexagonal layers[32]. Based on its full width at half maximum (FWHM), an estimate of the crystalline quality can be made. As a matter of fact, the FWHM of specific Raman bands is a key indicator of structural disorder and crystallite size[33]. Broader FWHM values typically reflect increased amorphization, reduced crystallite domains, or lattice distortions. The FWHM of the h-BN line in **Fig. 4(a)** is about 50 cm$^{-1}$ which indicates a moderate crystallinity since it is larger than the value corresponding to high crystalline h-BN (typically ~10 cm$^{-1}$ [32,34]) and close to that of multilayer or bulk-like h-BN [35,36]. This could be a consequence of a smaller substrate temperature used here compared to conventional h-BN deposition processes[1]. Furthermore, the FWHM of the h-BN line is about 2 times larger than in ref.[3] and slightly smaller than in ref.[2], both using ns-pulsed MHCDs with Al$_2$O$_3$ dielectric and larger (400 μm) MHCD holes. This value can be used to estimate a theoretical average crystallite size ($L_a$), as follows[33]:

$$L_a\left(cm^{-1}\right) = \frac{1417\left(cm^{-1}\right)}{\left[\Gamma_{1/2}\left(cm^{-1}\right) - 8.7\left(cm^{-1}\right)\right]} \quad (7)$$

where $\Gamma_{1/2}$ denotes the FWHM measured in **Fig. 4(a)**. This calculation gives $L_a \approx 3.5$ nm being slightly larger than in ref.[2] and about 2.5 times smaller than in ref.[3]. This small domain size could be attributed to a random nucleation of h-BN on the Si substrate[37], which can be related to plasma-induced instabilities among other factors, as implied by **Fig. 3(a)**. Furthermore, the raw Raman profile in **Fig. 4(a)** appears distorted on its left and right wings. In the former case, a broad, blue-shifted peak around 1350 cm$^{-1}$ (yellow line) is distinguishable with a FWHM=154 cm$^{-1}$. This may be indicative of the presence of amorphous carbon, a disorder-induced D–band or a residual component[1]. Jacquemin et al.[3] obtained a similar broad peak in the Raman spectrum of BN being, however, red-shifted. Its presence was attributed to the chemical instability of the film due to its exposure to moisture which induces a rapid hydrolysis (within 12–48 h) into boric acid (H$_3$BO$_3$). In the latter case, the right wing of the raw profile is affected by a weak peak around 1450 cm$^{-1}$ which exhibits a similar FWHM (=58 cm$^{-1}$) with the h-BN peak. This peak can be induced by surface impurities or deposit contamination. It is also remarkable that this peak coincides with a peak observed on the reference Raman signal on the Si substrate (light green in **Fig. 4(a)**), as observed in a previous work as well[2].

**Fig. 4(b)** shows representative SEM images of the same deposited h-BN film acquired at decreasing magnification (**Figs. 4(b1)–4(b6)**), enabling detailed analysis of surface morphology. The surface is covered with blister-like, nearly spherical features varying in size from below 100 nm (**Fig. 4(b1)**) up to over 100 μm (**Fig. 4(b6)**), revealing a hierarchical surface structure. In higher-magnification views (**Figs. 4(b1)** and **(b2)**) the h-BN film fully covers the substrate, with **Fig. 4(b2)** suggesting coalescence of individual blister domains, forming a textured and continuous layer. At intermediate magnification (**Figs. 4(b3)** and **(b4)**), localized defects are visible (outlined in yellow). These are not seen in regions shown in **Figs. 4(b5)** and **(b6)** denoting spatial inhomogeneity in film integrity. Notably, the defect in **Fig. 4(b3)** represents an uncovered area of the substrate, possibly the result of a ruptured blister followed by partial delamination due to surface tension gradients during growth. While incomplete film deposition in the areas covered by these defects is also possible, this is less probable in our system given the rather uniform discharge emission near the substrate (see zone Y$_1$ in **Fig. 3(a)**).

The former scenario is more likely to happen since the defects observed in **Fig. 4(b4)** contain wrinkle and blister remnants, indicating stress-induced mechanical failures.

Additional sources of film damage may include rapid cooling after synthesis or exposure to ambient humidity, both of which can induce thermal and mechanical stresses, leading to wrinkles, blisters or total delamination[38]. The large-area scan (**Fig. 4(b6)**) reveals non-uniform blister density but no evident defects, suggesting that defect formation is highly localized. Additional factors inducing this variation refer to the flow uniformity of the precursor, substrate pretreatment (e.g., plasma-based) and heating configuration. Particularly, the central heating of the substrate via a graphite resistor may induce radial temperature gradients, potentially leading to asymmetries in grain growth and defect formation. These findings will be considered to guide future improvements of the deposition reactor to achieve high uniformity and film quality.

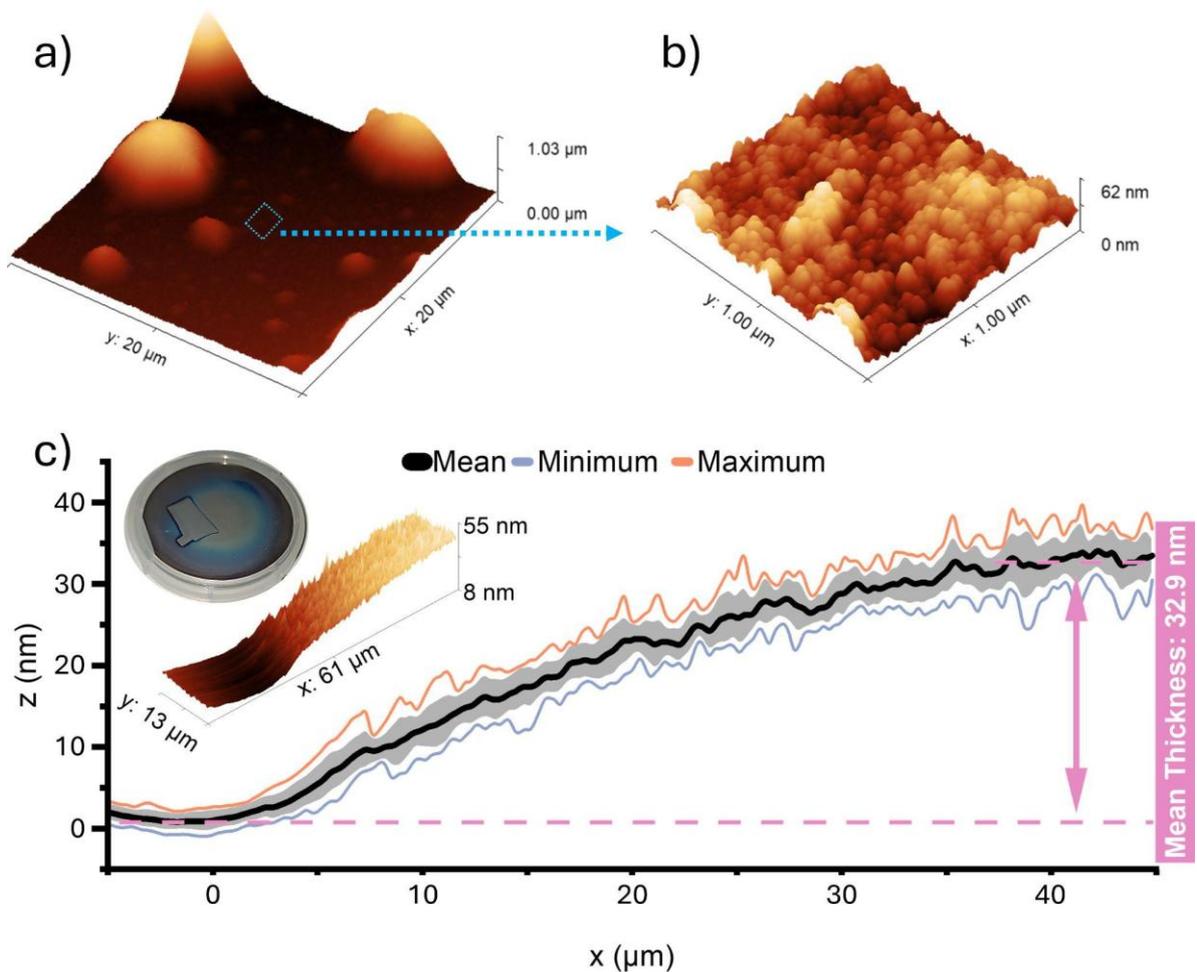

**FIG. 5**. AFM topography of h-BN on an area of the Si wafer of (a) 20×20 μm² and (b) 1×1 μm² (blue square in (a)); (c) h-BN film thickness obtained via AFM performed in two wafer zones: (i) without h-BN (achieved by placing a small Si specimen on the wafer and removing it after deposition, see inset photo; dashed pink line) and (ii) with h-BN (black line). The measurement is performed over a film length of 61 μm (see inset).

A quantitative analysis on the surface features of the h-BN films deposited is performed using AFM (**Fig. 5**). **Fig. 5(a)** shows a zoom over a 20×20 μm² region of the film with a maximum height of 1.03 μm, revealing an undulating pattern with blisters and flattened zones (similar to **Fig. 4(b)**). This image presents a mesoscale surface roughness which could be a consequence of temperature gradients on the substrate or plasma-induced inhomogeneities during the deposition. Both effects are probably due to the resistive heating method used as well as the non-uniform discharge pattern developed in the deposition chamber (Fig. 3(a)). Fig. 5(b) focuses on a 1×1 μm² region of the deposit portion shown in Fig. 5(a) (outlined with a blue dashed line). The maximum height of this zone is 62 nm, revealing a granular and rugged texture possibly comprising densely packed nanocrystalline grains and nucleation seeds. These figures shed light on important local properties of the deposit determining its quality, such as the root mean squared ($S_q$) and mean ($S_a$) roughness (related to surface smoothness), maximum peak height ($S_p$) (indicative of blisters, wrinkles, etc.), skew ($S_{sk}$) (showing a deviation from a smooth surface) and excess kurtosis ($S_{ek}$) (defining sharp or flat features). These have been measured using the Gwyddion software and are listed in **Tab.1**.

TAB. 1. Surface quantities of the h-BN film measured with AFM

| *Quantity* | Rms Roughness ($S_q$) | Mean Roughness ($S_a$) | Peak height ($S_p$) | Skewness ($S_{sk}$) | Kurtosis ($S_{ek}$) |
|---|---|---|---|---|---|
| Fig. 5(a) | 139.4 μm | 91.2 nm | 0.855 μm | 2.266 | 5.614 |
| Fig. 5(b) | 9.1 nm | 7 nm | 36.65 nm | 0.59 | 0.62 |

Figs. 5(a) and (b) reveal important variations in the film morphology over different scales. From Tab. 1 it can be concluded that the film region in Fig. 5(a) exhibits quite large values of $S_q$, $S_a$ and $S_p$ as well as a strong positive skewness and kurtosis. This is indicative of dominant sharp protrusions on the surface possibly occurring by a multilayer aggregation of h-BN. However, the film's portion analysed in Fig. 5(b) appears much smoother with a relatively symmetric height distribution pointing to a more uniform grain-scale deposition. These discrepancies on the film topography are in agreement with the Raman results suggesting a moderate crystallinity and structural disorder. Finally, the thickness of the deposited film in this scale can be measured with AFM. To achieve this, an area of a new Si wafer was blocked by a second small rectangular Si specimen (see inset in Fig. 5(c)). This helped in maintaining a region of the new wafer unexposed to plasma, thus avoiding coverage with h-BN. Then, the small Si specimen was removed after deposition and the new wafer was analysed via AFM. Analysis was performed starting from an uncovered region of the wafer and moving towards the h-BN-covered zone. The total zone scanned refers to an area of 13×61 μm² (shown in the inset of Fig. 5(c)) and is chosen to be as flat as possible (i.e., avoiding blisters/wrinkles). From this analysis, an average film thickness of about 33 nm is found (Fig. 5(c)). This corresponds to a deposition rate of about 22 nm/h which is ≈1.3 times smaller than in ref.[2] and more than 2–fold larger compared to ref.[3]. Given that monolayer h-BN has a thickness of 0.33 nm[39,40], the measured thickness in the present study corresponds to about 100 atomic layers being representative of a thin multilayer film. This aligns with the measured Raman FWHM of h-BN deposited indicating moderate crystallinity possibly due to turbostratic stacking or other effects.

To investigate the elemental composition and lateral distribution of elements across the deposited film, XPS measurements were performed along the sample's diameter (800 μm resolution; yellow dashed line in Fig. 6(a)) and within a central 9 mm² area (90 μm resolution; yellow dashed square in Fig. 6(a)). Fig. 6 presents the chemical characterization and spatial compositional analysis of the grown h-BN. A representative XPS spectrum is shown in Fig. 6(a) with the analyzed sample inset. The spectrum displays distinct peaks of B1s (191 eV), C1s (285 eV), N1s (398 eV) and O1s (533 eV), confirming their presence in the film. The variation in atomic concentrations and the B/N atomic ratio along the wafer diameter are shown in Fig. 6(b). The small content of carbon (8.5%±1%) detected probably originates from the graphite substrate heater and/or the PTFE mount used to secure the MHCD structure between the two chambers. Additionally, the unexpected detection of atomic oxygen (14%±2%) is possibly due to post-deposition oxidation when the wafer is exposed to ambient air[2,38].

Although oxygen incorporation could also arise from the dielectric material in the MHCD device, this seems impossible here, as aluminum nitride (AlN) was used instead of oxygen-rich alumina ($Al_2O_3$) as dielectric material. Other possible contamination sources include trace impurities of oxygen in the process gases. Further investigation is needed to limit its presence in the film.

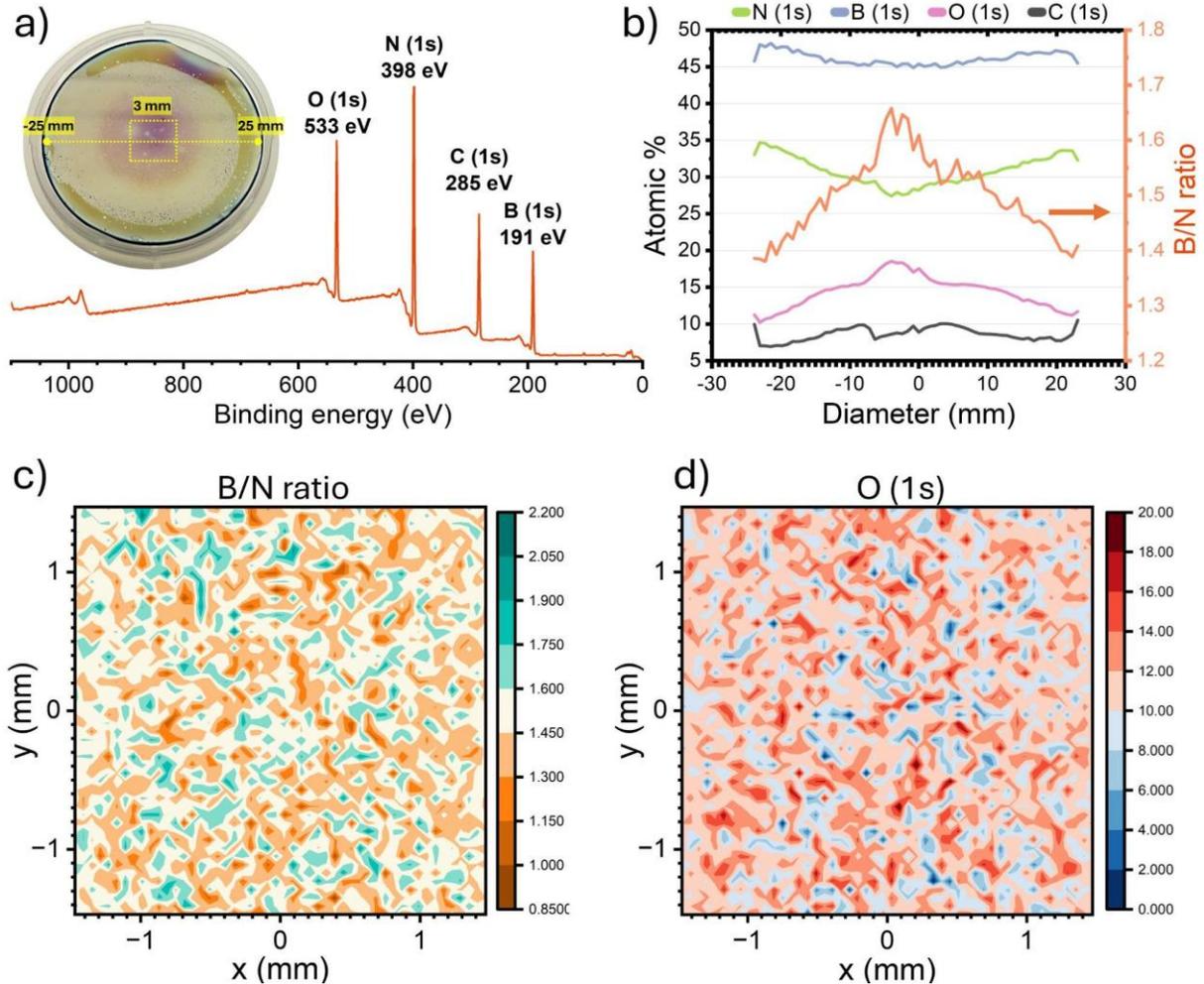

**FIG. 6**. XPS analysis of the grown h-BN film on a Si<100> wafer (5 cm diameter; inset in (a)). (a) Indicative XPS spectrum displaying characteristic peaks for B(1s), N(1s), O(1s) and C(1s) detected. (b) Atomic concentration profiles and B/N ratio extracted across the diameter of the wafer (yellow dashed line in (a)). (c) 2D map of the B/N ratio over the yellow square shown in (a). (d) 2D distribution of O(1s) concentration (%) in the same region with (c).

Furthermore, B and N atoms are the most abundant elements detected across the diameter of the film. This result, in combination with the Raman signature of h-BN identified (**Fig. 4(a)**), confirms the potential of MHCD technology for synthesizing h-BN films over large surfaces (here ≈20 $cm^2$). The atomic percentage of boron (46±1%) dominates and remains fairly constant across the wafer, while that of nitrogen exhibits a small dip close to the center (≈30%) and increases towards the edges (≈35%), resulting in an average nitrogen content of 31±2%. These variations affect the B/N atomic ratio ($R_{B/N}$), which appears spatially correlated with morphological irregularities observed previously such as structural disorder and surface wrinkles. It is noteworthy to mention that the $R_{B/N}$ profile resembles the discharge emission pattern labelled $Y_2$ in **Fig. 3(a)**. Since the shape of $R_{B/N}$ is largely determined by that of the nitrogen signal, the resemblance implies a possible link between the discharge morphology and film composition. However, $Y_2$ is relatively far from the substrate, while the discharge zone found near the substrate ($Y_1$) is more likely to affect film growth, its emission profile appearing more uniform. Further investigation is necessary to resolve this, e.g., in-situ mapping of plasma-generated atomic nitrogen density via laser-based diagnostics, as done in small portable MHCD reactors[9,12]. However, applying such diagnostics to this deposition reactor remains a challenge due to reactor size and access limitations.

Despite these ambiguities, the average $R_{B/N}$ across the sample in **Fig. 6(b)** ($\approx$1.5) remains relatively close to the ideal stoichiometry of h-BN (i.e., $R_{B/N}$=1). Slight deviations may arise from carbon and/or oxygen incorporation in the film. These could be affected by a variation of the deposition time, MHCD geometry, as well as the distance between the MHCD and the substrate[2,3]. Other process parameters such as gas mixture composition and purity, precursor flow rate, chamber pressure, substrate heating and properties of the HV power supply should be explored to better control stoichiometry and improve film uniformity. Optimizing these factors will be crucial for reliably producing high-quality h-BN via the MHCD method.

**Figs. 6(c)** and **(d)** show 2D compositional maps of the $R_{B/N}$ and O(1s) signal intensity over a 9 mm$^2$ area centered on the substrate. These maps yield average values of $R_{B/N}$=1.5$\pm$0.05% and O content =11$\pm$1%, both consistent with the scans performed along the diameter (**Figs. 6(b)**). Across the mapped area, the $R_{B/N}$ remains relatively close to the stoichiometric unity, although noticeable localized deviations are evident. These may be attributed to non-uniform surface coverage, as observed in SEM (**Figs. 4(b)**), or to nanoscale imperfections revealed by AFM (**Figs. 5(a)** and **(b)**). Oxygen incorporation shown in **Fig. 6(d)** could be more concentrated near areas associated with morphological defects, supporting the hypothesis that post-deposition oxidation (due to ambient exposure or thermal stress) contributes to localized surface chemistry variation. Overall, these compositional patterns imply that the inhomogeneities are more likely a consequence of downstream mechanical or thermal effects.

## IV. CONCLUSION

We used a ns-pulsed N$_2$/Ar Micro Hollow Cathode Discharge (MHCD) as a microplasma source to deposit hexagonal boron nitride (h-BN) on Si <100> substrate ($\approx$20 cm$^2$) via Plasma Enhanced Chemical Vapor Deposition (PECVD). This MHCD uses for the first time aluminum nitride (AlN) as dielectric material to suppress film contamination by atomic oxygen as much as possible, an issue previously observed with widely employed Al$_2$O$_3$ dielectrics. The deposited h-BN was characterized using multiple diagnostics, including Raman spectroscopy, scanning electron microscopy (SEM), atomic force microscopy (AFM), and X-ray photoelectron microscopy (XPS). The plasma properties (emission profiles, gas temperature, discharge morphology) were probed *in-situ* during h-BN deposition using optical emission spectroscopy (OES) and ICCD imaging.

This study validates the use of AlN as dielectric material in ns-pulsed MHCD for the synthesis of h-BN over large-surface Si<100> substrates. The successful deposition is demonstrated by the characteristic E$_{2g}$ Raman mode around 1367 cm$^{-1}$, widespread substrate surface coverage revealed by SEM, and fingerprints of boron and nitrogen atoms in the film identified by XPS. Replacing classic Al$_2$O$_3$ with AlN excluded the dielectric material itself as a potential source of film oxidation. Unfortunately, a small percentage of atomic oxygen is still present in the films most probably due to post-deposition oxidation (due to ambient exposure or thermal stress). SEM and AFM imaging show localized defects, blisters, and delamination, likely induced by thermal gradients, rapid cooling (~13 K/min), and ambient exposure. Furthermore, AFM analysis indicates an average film thickness of about 33 nm after 90 minutes of deposition (~22 nm/h deposition rate). XPS reveals an average B/N atomic ratio of ~1.5 over a 9 mm$^2$ central area and along the substrate diameter (5 cm), with localized deviations spatially correlated with surface imperfections.

OES allows identifying the main reactive species generated and estimating the gas temperature, being smaller than that of the substrate. Furthermore, it allows assessing important mechanisms driving the generation of atomic nitrogen, one of the two key species for the film synthesis. ICCD measurements reveal non-uniform spatial discharge properties in the bulk and quasi-uniform plasma emission close to the substrate. These gradients may also contribute to the compositional and structural inhomogeneities observed in the films. These results highlight the need for improved thermal management, optimized MHCD geometry, and better control of plasma-substrate interactions. Further implementation of advanced *in-situ* diagnostics (e.g., laser-based reactive species mapping) may enable real-time feedback for reactor design and process refinement.

## ACKNOWLEDGMENTS

This work was supported by the French Research National Agency (ANR) via the project SPECTRON (ANR-23-CE51-0004-01), PlasBoNG (ANR-20-CE09-0003-01), DESYNIB (ANR-16-CE08-0004), and LabEx SEAM (ANR-10-LABX-0096 and ANR-18-IDEX-0001). The authors would like to thank Laurent Invernizzi for his assistance with the installation of optical diagnostics, Ludovic William, Nicolas Fagnon, and Noel Girodon-Boulandet for technical drawing and support, Maria Konstantakopoulou and Sarah Al Zeibak for their technical support with the SEM analyses, Hamza Bouhriz for assistance on the AFM measurements, and Pascal Martin for support with the XPS measurements.

## AUTHOR DECLARATIONS

### Conflict of Interest

The authors have no conflicts to disclose.

### Author Contributions

**Belkacem Menacer**: Data curation (lead); Formal analysis (equal); Investigation (lead); Methodology (supporting); Visualization (supporting); Writing–original draft (equal); Writing–review & editing (supporting). **Dimitrios Stefas**: Data curation (equal); Visualization (lead); Formal analysis (equal); Writing–original draft (equal); Writing–review & editing (equal); Validation (equal). **Nikolaos Chazapis**: Investigation (supporting); Validation (supporting); Writing–review & editing (supporting). **Claudia Lazzaroni**: Conceptualization (equal); Funding acquisition (lead); Methodology (equal); Project administration (lead); Supervision (lead); Validation (supporting); Writing–review & editing (supporting); Resources (lead). **Kristaq Gazeli**: Conceptualization (equal); Formal analysis (equal); Funding acquisition (lead); Methodology (equal); Project administration (lead); Supervision (lead); Writing–original draft (lead); Writing–review & editing (lead); Resources (lead). **Vianney Mille**: Conceptualization (equal); Funding acquisition (equal); Methodology (equal); Project administration (equal); Supervision (lead); Validation (supporting); Writing–review & editing (supporting).

## DATA AVAILABILITY

The data that support the findings of this study are available from the corresponding author upon request.

## REFERENCES

[1] Y.-J. Cho, A. Summerfield, A. Davies, T.S. Cheng, E.F. Smith, C.J. Mellor, A.N. Khlobystov, C.T. Foxon, L. Eaves, P.H. Beton, and S.V. Novikov, "Hexagonal Boron Nitride Tunnel Barriers Grown on Graphite by High Temperature Molecular Beam Epitaxy," Sci. Rep. **6**(1), 34474 (2016).
[2] H. Kabbara, S. Kasri, O. Brinza, G. Bauville, K. Gazeli, J. Santos Sousa, V. Mille, A. Tallaire, G.

Lombardi, and C. Lazzaroni, "A microplasma process for hexagonal boron nitride thin film synthesis," Appl. Phys. Lett. **116**(17), 171902 (2020).

[3] M. Jacquemin, A. Remigy, B. Menacer, V. Mille, C. Barraud, and C. Lazzaroni, "Amorphous and hexagonal boron nitride growth using bromide chemistry," MRS Commun. **14**(1), 63–68 (2024).

[4] J. Wang, F. Ma, W. Liang, and M. Sun, "Electrical properties and applications of graphene, hexagonal boron nitride (h-BN), and graphene/h-BN heterostructures," Mater. Today Phys. **2**, 6–34 (2017).

[5] S.-J. Liang, B. Cheng, X. Cui, and F. Miao, "Van der Waals Heterostructures for High-Performance Device Applications: Challenges and Opportunities," Adv. Mater. **32**(27), 1903800 (2020).

[6] L. Britnell, R.V. Gorbachev, R. Jalil, B.D. Belle, F. Schedin, A. Mishchenko, T. Georgiou, M.I. Katsnelson, L. Eaves, S.V. Morozov, N.M.R. Peres, J. Leist, A.K. Geim, K.S. Novoselov, and L.A. Ponomarenko, "Field-Effect Tunneling Transistor Based on Vertical Graphene Heterostructures," Science **335**(6071), 947–950 (2012).

[7] V. Martin, G. Bauville, M. Fleury, and V. Puech, "Exciplex emission induced by nanosecond-pulsed microdischarge arrays operating at high repetition rate frequency," Plasma Sources Sci. Technol. **21**(6), 065001 (2012).

[8] S. Kasri, L. William, X. Aubert, G. Lombardi, A. Tallaire, J. Achard, C. Lazzaroni, G. Bauville, M. Fleury, K. Gazeli, S. Pasquiers, and J.S. Sousa, "Experimental characterization of a ns-pulsed micro-hollow cathode discharge (MHCD) array in a $N_2$/Ar mixture," Plasma Sources Sci. Technol. **28**(3), 035003 (2019).

[9] A. Remigy, S. Kasri, T. Darny, H. Kabbara, L. William, G. Bauville, K. Gazeli, S. Pasquiers, J. Santos Sousa, N. De Oliveira, N. Sadeghi, G. Lombardi, and C. Lazzaroni, "Cross-comparison of diagnostic and 0D modeling of a micro-hollow cathode discharge in the stationary regime in an Ar/$N_2$ gas mixture," J. Phys. Appl. Phys. **55**(10), 105202 (2021).

[10] A. Remigy, B. Menacer, K. Kourtzanidis, O. Gazeli, K. Gazeli, G. Lombardi, and C. Lazzaroni, "Absolute atomic nitrogen density spatial mapping in three MHCD configurations," Plasma Sources Sci. Technol. **33**(2), 025013 (2024).

[11] C. Lazzaroni, P. Chabert, A. Rousseau, and N. Sadeghi, "Sheath and electron density dynamics in the normal and self-pulsing regime of a micro hollow cathode discharge in argon gas," Eur. Phys. J. D **60**(3), 555–563 (2010).

[12] A. Remigy, X. Aubert, S. Prasanna, K. Gazeli, L. Invernizzi, G. Lombardi, and C. Lazzaroni, "Absolute N-atom density measurement in an Ar/$N_2$ micro-hollow cathode discharge jet by means of ns-two-photon absorption laser-induced fluorescence," Phys. Plasmas **29**(11), 113508 (2022).

[13] K.H. Schoenbach, R. Verhappen, T. Tessnow, F.E. Peterkin, and W.W. Byszewski, "Microhollow cathode discharges," Appl. Phys. Lett. **68**(1), 13–15 (1996).

[14] B. Du, M. Aramaki, S. Mohr, Y. Celik, D. Luggenhölscher, and U. Czarnetzki, "Spatially and temporally resolved optical spectroscopic investigations inside a self-pulsing micro thin-cathode discharge," J. Phys. Appl. Phys. **44**(25), 252001 (2011).

[15] C. Lazzaroni, P. Chabert, A. Rousseau, and N. Sadeghi, "The excitation structure in a micro-hollow cathode discharge in the normal regime at medium argon pressure," J. Phys. Appl. Phys. **43**(12), 124008 (2010).

[16] H. Wei, N. Wang, Z. Duan, and F. He, "Experimental and simulation study of pulsed micro-hollow cathode discharge in atmospheric-pressure helium," Phys. Plasmas **25**(12), 123513 (2018).

[17] Z. Duan, P. Li, F. He, R. Han, and J. Ouyang, "Study on the characteristics of helium plasma jet by pulsed micro-hollow cathode discharge," Plasma Sources Sci. Technol. **30**(2), 025001 (2021).

[18] S. He, J. Deng, Y. Qiao, Q. Li, and L. Dong, "Experiment and simulation on the micro-hollow cathode sustained discharge in helium with different geometries of the second anode," J. Appl. Phys. **133**(8), 083301 (2023).

[19] D.L. Rusterholtz, D.A. Lacoste, G.D. Stancu, D.Z. Pai, and C.O. Laux, "Ultrafast heating and oxygen dissociation in atmospheric pressure air by nanosecond repetitively pulsed discharges," J. Phys. Appl. Phys. **46**(46), 464010 (2013).

[20] P.J. Bruggeman, N. Sadeghi, D.C. Schram, and V. Linss, "Gas temperature determination from rotational lines in non-equilibrium plasmas: a review," Plasma Sources Sci. Technol. **23**(2), 023001 (2014).

[21] T.D. Nguyen, and N. Sadeghi, "Rotational and vibrational distributions of $N_2$(C $3\Pi_u$) excited by

state-selected Ar($^3P_2$) and Ar($^3P_0$) metastable atoms," Chem. Phys. **79**(1), 41–55 (1983).
[22] M. Tabbal, M. Kazopoulo, T. Christidis, and S. Isber, "Enhancement of the molecular nitrogen dissociation levels by argon dilution in surface-wave-sustained plasmas," Appl. Phys. Lett. **78**(15), 2131–2133 (2001).
[23] A. Bogaerts, "Hybrid Monte Carlo — Fluid model for studying the effects of nitrogen addition to argon glow discharges," Spectrochim. Acta Part B At. Spectrosc. **64**(2), 126–140 (2009).
[24] N. Sadeghi, D.W. Setser, A. Francis, U. Czarnetzki, and H.F. Döbele, "Quenching rate constants for reactions of Ar(4p′[1/2], 4p[1/2], 4p[3/2]$_2$, and 4p[5/2]$_2$) atoms with 22 reagent gases," J. Chem. Phys. **115**(7), 3144–3154 (2001).
[25] J.E. Velazco, J.H. Kolts, and D.W. Setser, "Rate constants and quenching mechanisms for the metastable states of argon, krypton, and xenon," J. Chem. Phys. **69**(10), 4357–4373 (1978).
[26] R.P. Cardoso, T. Belmonte, P. Keravec, F. Kosior, and G. Henrion, "Influence of impurities on the temperature of an atmospheric helium plasma in microwave resonant cavity," J. Phys. Appl. Phys. **40**(5), 1394 (2007).
[27] N. Kang, F. Gaboriau, S. Oh, and A. Ricard, "Modeling and experimental study of molecular nitrogen dissociation in an Ar–N$_2$ ICP discharge," Plasma Sources Sci. Technol. **20**(4), 045015 (2011).
[28] S. Mazouffre, C. Foissac, P. Supiot, P. Vankan, R. Engeln, D.C. Schram, and N. Sadeghi, "Density and temperature of N atoms in the afterglow of a microwave discharge measured by a two-photon laser-induced fluorescence technique," Plasma Sources Sci. Technol. **10**(2), 168 (2001).
[29] A. Salmon, N.A. Popov, G.D. Stancu, and C.O. Laux, "Quenching rate of N($^2$P) atoms in a nitrogen afterglow at atmospheric pressure," J. Phys. Appl. Phys. **51**(31), 314001 (2018).
[30] P.A. Sá, and J. Loureiro, "A time-dependent analysis of the nitrogen afterglow in and - Ar microwave discharges," J. Phys. Appl. Phys. **30**(16), 2320 (1997).
[31] A. Ricard, J. Tetreault, and J. Hubert, "Nitrogen atom recombination in high pressure Ar-N$_2$ flowing post-discharges," J. Phys. B At. Mol. Opt. Phys. **24**(5), 1115 (1991).
[32] R.V. Gorbachev, I. Riaz, R.R. Nair, R. Jalil, L. Britnell, B.D. Belle, E.W. Hill, K.S. Novoselov, K. Watanabe, T. Taniguchi, A.K. Geim, and P. Blake, "Hunting for Monolayer Boron Nitride: Optical and Raman Signatures," Small **7**(4), 465–468 (2011).
[33] R.J. Nemanich, S.A. Solin, and R.M. Martin, "Light scattering study of boron nitride microcrystals," Phys. Rev. B **23**(12), 6348–6356 (1981).
[34] H.X. Jiang, and J.Y. Lin, "Hexagonal boron nitride for deep ultraviolet photonic devices," Semicond. Sci. Technol. **29**(8), 084003 (2014).
[35] S. Saha, A. Rice, A. Ghosh, S.M.N. Hasan, W. You, T. Ma, A. Hunter, L.J. Bissell, R. Bedford, M. Crawford, and S. Arafin, "Comprehensive characterization and analysis of hexagonal boron nitride on sapphire," AIP Adv. **11**(5), 055008 (2021).
[36] A.K. Dąbrowska, M. Tokarczyk, G. Kowalski, J. Binder, R. Bożek, J. Borysiuk, R. Stępniewski, and A. Wysmołek, "Two stage epitaxial growth of wafer-size multilayer h-BN by metal-organic vapor phase epitaxy – a homoepitaxial approach," 2D Mater. **8**(1), 015017 (2020).
[37] R.Y. Tay, S.H. Tsang, M. Loeblein, W.L. Chow, G.C. Loh, J.W. Toh, S.L. Ang, and E.H.T. Teo, "Direct growth of nanocrystalline hexagonal boron nitride films on dielectric substrates," Appl. Phys. Lett. **106**(10), 101901 (2015).
[38] J. Möller, D. Reiche, M. Bobeth, and W. Pompe, "Observation of boron nitride thin film delamination due to humidity," Surf. Coat. Technol. **150**(1), 8–14 (2002).
[39] H. Park, G.H. Shin, K.J. Lee, and S.-Y. Choi, "Atomic-scale etching of hexagonal boron nitride for device integration based on two-dimensional materials," Nanoscale **10**(32), 15205–15212 (2018).
[40] S. Wanigasekara, K. Rijal, F. Rudayni, M. Panth, A. Shultz, J.Z. Wu, and W.-L. Chan, "Using an Atomically Thin Layer of Hexagonal Boron Nitride to Separate Bound Charge-Transfer Excitons at Organic Interfaces," Phys. Rev. Appl. **18**(1), 014042 (2022).